\documentclass{aa}
\usepackage{epsf,epsfig}

\def\refe{}  

\begin{document}

\title{Reviving Dead Zones in Accretion Disks \\ by Rossby Vortices at their Boundaries}

\author{P. Varni\`ere \inst{1,2} \&  M. Tagger \inst{3}}
\institute{ Department of Physics \& Astronomy, Rochester University, 
Rochester NY 14627-0171 
\and
 LAOG, Universit\'e J. Fourier UMR $5571$ France, \email{pvarni@pas.rochester.edu}
\and
 Service d'Astrophysique (UMR Astroparticules et cosmologie) CEA/Saclay F-91191 Gif sur Yvette 
 \email{tagger@cea.fr}}
 
\abstract{
 Models of the accretion disks of Young Stellar Objects show that they 
should not be ionized at a few AU from the star, and thus not subject to the 
MHD turbulence believed to cause accretion. This has been suggested to create a 
'Dead Zone' where accretion remains unexplained. Here we show that the existence 
of the Dead Zone  self-consistently creates a density profile favorable to
the Rossby Wave Instability of Lovelace et al. (1999). This
instability will create and sustain Rossby vortices in the disk which 
could lead to enhanced planet formation.
}
  \maketitle

\section{Introduction}
    The existence of a Dead Zone in the  accretion disk of  Young Stellar Objects (YSO) 
    was first proposed
    by \cite{G96}; he found that  the inner region of the disk is ionized by collisions, 
    and the outer region by cosmic rays. On the other hand there exists an intermediate region, 
    dubbed a Dead Zone, where the ionization remains so low that the gas is not coupled to 
    magnetic fields. This creates a problem since accretion is widely believed to result from MHD 
    instabilities, such as the Magneto-Rotational Instability (\cite{BH91}). The Dead Zone should thus 
    inhibit accretion in a significant portion of the disk. 
   More recent studies (for example \cite{A98,R01,MP05}) have put 
    constraints on the position and extent of the Dead Zone depending on the
    heating mechanisms and disk properties, and related them to observations. Various estimates have 
    been given for the extent of the Dead Zone, which is typically found to range between 1 and 5 AU from the star.

 In the present paper we discuss the consequences on the disk dynamics of the presence of 
 such a zone, where the transport of matter and angular momentum (which we will represent by a turbulent viscosity) 
 is significantly lower than elsewhere in the disk. 
    In particular we show that an important role could be played by the Rossby Wave Instability 
    (RWI), introduced by Lovelace {\em et al.} (1999) as a possible  
    means of transporting angular momentum in non-magnetized disks. This  
    instability requires locally an extremum of a certain quantity, which  
    we will discuss below. 
    
    Here we show that the existence of a   
    Dead Zone naturally generates such an extremum at its boundaries. The RWI naturally generates Rossby vortices. We relate this to the observation by several authors (\cite{BS95,Ch00,JAB04}), that vortices can speed up  
    planet formation, by allowing planetesimals to accumulate in their  
   center without following the viscous accretion. 
   As shown in \cite{BS95}, this might solve a  
   problem encountered in present views of planet formation, where  
   planetesimals accrete to the central star on a viscous time scale,  
   before they have had time to form a planet.
   We also show that the instability extends across the Dead Zone as spiral waves, which can maintain  
   a significant accretion rate despite the absence of locally generated turbulence.
\section{Rossby Waves Instability}
    In 1999 Lovelace {\em et al.} (see also Li {\em et al.} 2000, 2001), elaborating on
    previous work by \cite{L78}, found a non-axisymmetric instability  
    (Rossby Wave Instability, or RWI) occurring in unmagnetized disks, 
   when there is a local extremum in the radial profile of a quantity ${\cal L}$.
\begin{eqnarray}
{\cal L} =  {\cal F}(r) S^{2/\Gamma}(r) = \frac{\Sigma \Omega}{\kappa^{2}} 
                    \frac{p}{\Sigma^{\Gamma}}
\end{eqnarray}
Where $\Sigma$ is the surface density, $\Omega$ the rotation frequency, $\kappa$ the
     epicyclic frequency and $\Gamma$ is the adiabatic index. With $\Gamma =1$, ${\cal L}$
     reduces to {\refe the inverse of} the vortensity responsible for the corotation resonance of the Papaloizou-Pringle instability
     (\cite{PP85,GGN86,PL89}).
Numerical simulations by \cite{L00} confirmed  that a bump or jump in the surface density $\Sigma$ or the pressure $p$ can  cause the instability, and studied its consequences. Here we show how such a bump in the radial density profile can naturally result from the presence of a Dead Zone in the disk of a young stellar object (YSO).
\section{Numerical Model of the Dead Zone/ Instability Criterion}
\subsection{Modelling the Dead Zone}
The Dead Zone is a  limited region in the disk where the
    gas is not well coupled with the magnetic field, because  
    the ionization is too low. If we make the assumption, as in 
    \cite{G96}, that turbulent 
    accretion is due to MHD instabilities such as the Magneto-Rotationnal
    Instability (MRI, \cite{BH91}), we can model that region by a drop
    of the turbulent (MRI-generated) transport.

    We represent accretion in the disk by a turbulent  viscosity 
    (\cite{SS73}) which drops in the region of the dead zone, where  
   MHD instabilities are inactive.
    We include this viscosity profile in the $2$D non-linear hydrodynamic code
    developed by \cite{masset02,masset03}. This is an Eulerian
    polar grid code with a staggered mesh and an artificial second
    order viscous pressure to stabilize the shocks (see also
    \cite{stone}).  The code uses the FARGO algorithm which significantly 
    accelerates computations in a differentially rotating disk.
    {\refe The simulation are done using standard boundary conditions, 
    namely accretion is allowed through the inner boundary by using open 
    boundary conditions and we assume the disk to be larger by allowing mass 
    to flow in from the outer boundary. An axisymmetric, 
    steady-state solution, which implies $\nu\ \Sigma=$ Const., would give 
    an artificially huge overdensity in the Dead Zone. Since the instability 
    would cancel this condition, we rather start from a simple power-law density 
    profile that is allowed to evolve self-consistently during the simulations. 
    Reaching a true steady state in these conditions would require prohibitive 
    computation time and we rather concentrate in this paper on the development 
    of the instability and a discussion of its consequences.}

    The runs we present here are done with a resolution of
    $N_r = 150$ and $N_\theta = 450$ grid points. 
    We performed two types of simulations: in the first one, 
    which allows only a limited resolution, the physical size of the grid 
    spans radii between $.25$ AU and $10$ AU, and the Dead Zone extends 
    from 1 to 5 AU. In the second one, for higher resolution, and since 
    we are mainly interested in what happens at the boundaries of the 
    Dead Zone, we use a grid spanning from $.25$ to $4$ AU,  with a 
    Dead Zone extending from .5 to 1 AU.

In both cases the disk aspect ratio $H/r = 0.04$ where
    $H$, the vertical scale height of the disk, is uniform and constant.
    The sound speed of the gas is set from the disk aspect ratio.
   The initial density profile is 
   $\Sigma(r) = 10^{-5} \left({r / R_{min}}\right)^{-1}$ where $R_{min}$
    is the radius of the inner edge of the grid. 
    Matter is allowed to flow in from the outer boundary,  by accretion from the 
    outer regions of the disk not included in the simulation.

\begin{figure}[htbp]
\centering
\resizebox{\hsize}{!}{\includegraphics{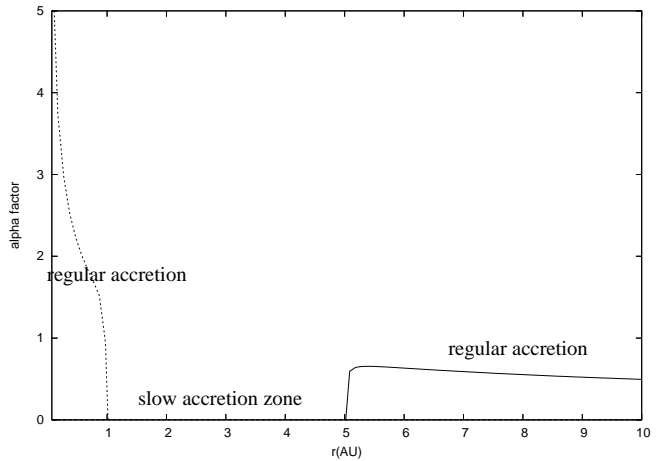}}
   \caption{Profile of the $\alpha$-viscosity implemented to represent a Dead Zone between $1$ and $5$ AU
   with $(\epsilon, \delta_{r}) = (10^{-5}, 50)$.}
\label{fig:viscosity_profile}
\end{figure}

   In order to model the Dead Zone we use an arctangent drop in the radial viscosity
   profile: 
\begin{eqnarray}   
\left\{
\begin{array}{lll}
\nu = \nu_o \left( \epsilon + {\rm atan} (\delta_{r}(r -r^{DZ}_{out}))\right) &\mbox{if}  &r< r^{DZ}_{out}\nonumber\\
\nu = \nu_o \epsilon  &\mbox{if}  &r^{DZ}_{out}>r>r^{DZ}_{in}\nonumber\\
\nu = \nu_o \left( \epsilon + {\rm atan} (\delta_{r}(-r +r^{DZ}_{in}))\right) &\mbox{if}&  r<r^{DZ}_{in}\nonumber
\end{array}
\right.
\end{eqnarray}
where $\epsilon$ represents a residual viscosity in the Dead Zone, and $\delta_{r}$ defines the steepness of the viscosity profile at the edges of the
   Dead Zone. All the runs presented here are done with $\epsilon$ and $\delta_{r}$ equal to
   $10^{-5}$ and $50$ respectively, giving the profiles shown on figure \ref{fig:viscosity_profile}.
   {\refe We did runs with $\epsilon$ varying from $10^{-2}$  to $10^{-5}$ to confirm the behaviour
   with a less steep Dead-Zone.}
\subsection{Evolution of the RWI Criterion}
\begin{figure}[htbp]
\centering
\epsfig{file=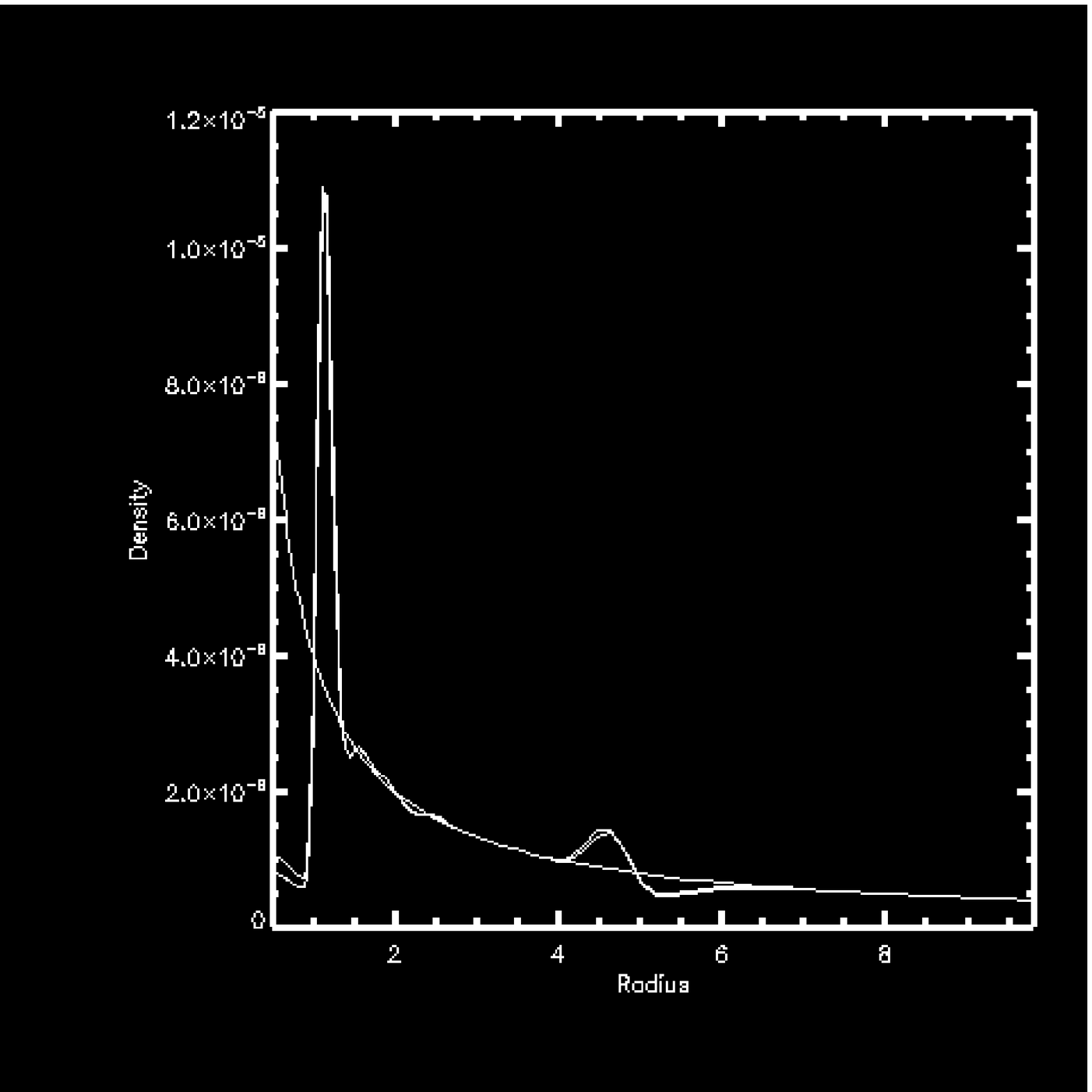,width=4.5cm}
\epsfig{file=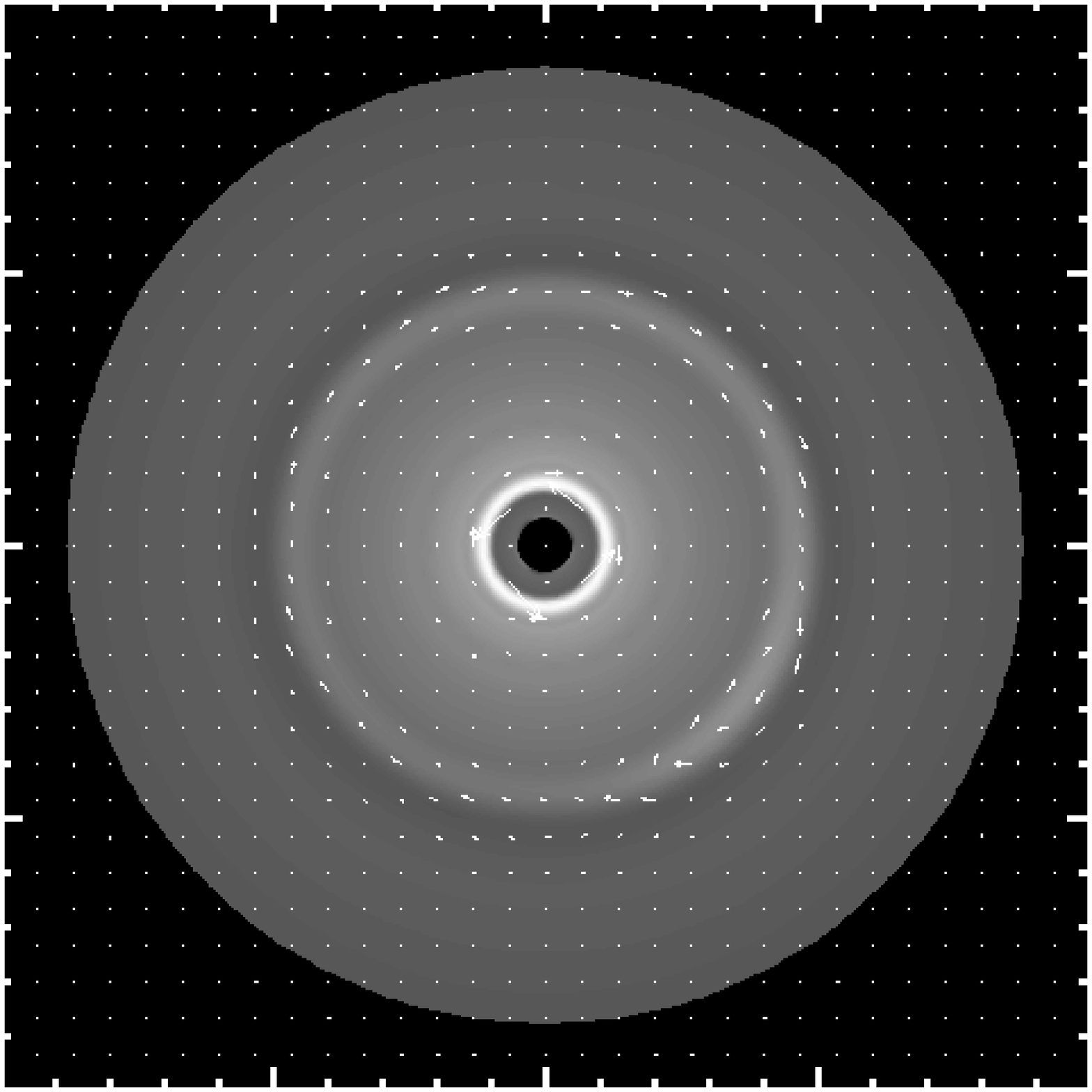,width=4.5cm}
\caption{Left: plot of $\cal L$ a $t=0$ and once the vortices have formed.
right: plot of the density in the disk at t= 100 years, with the velocity field superimposed. The velocity field shows the vortex at $5$ AU. 
There is also one at $1$AU, not visible here.}
\label{fig:criteria_RWI_out1}
\end{figure}

   With the prescribed viscosity profile, viscous accretion occurs in the outer region 
   of the simulation but is far slower in the Dead Zone. 
   {\refe As a result,
   gas accumulates near the outer edge of this zone}, and naturally produces in less
   than a hundred orbits at that radius the extremum of $\cal L$  required by the RWI
    (see fig \ref{fig:criteria_RWI_out1} of the low resolution run)
   {\refe If we now look at the inner edge of the Dead Zone, the opposite happens,
   namely the matter in the Dead-Zone is accreted slower than that in the inner 
   region, therefore creating a dip inward from the Dead-Zone, also producing
   a favorable profile of $\cal L$}
   
\subsection{Formation of Vortices}
Once the threshold for the instability is reached we see Rossby vortices developping in the disk
    (there was still enough of the initial random perturbation to seed the instability). 

\begin{figure}[htbp]
\centering
\resizebox{\hsize}{!}{\includegraphics{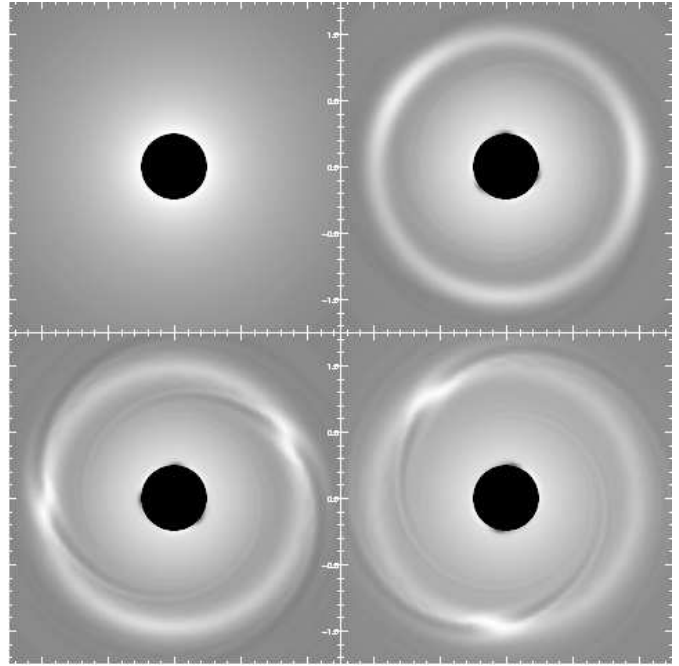}}
   \caption{Zoom of the first 2 inner AU of the simulation at $t=0, 100, 200, 300$ years,
   showing the density. One sees three vortices forming, later evolving to two vortices, near the  
   outer edge of the Dead Zone.}
\label{fig:disk_RWI_out1}
\end{figure}

Figure \ref{fig:disk_RWI_out1} shows a surface density plot of the disk at 
     $t=0,100,200,300$ years. We first see three vortices forming, but later only two of them surviving. 
These vortices are anticyclonic, and it is known (\cite{BS95, Ch00, FMB01,JAB04}) that this can cause an accumulation of dust and enhanced formation of planetesimals.
\cite{BS95}, computed the amount of mass captured by a vortex depending on its distance to the star. Using our Dead Zone extending from $1$ to $5$ AU we see that the vortex at the inner edge of the Dead Zone
     will be able to capture $0.6\ M_{\oplus}$ in $500$ years (= $500$ orbits). From the same calculation the vortex at the
     outer edge of the Dead Zone will accumulate $16\ M_{\oplus}$ in $6\ 10^{3}$ years (= $500$ orbits). Those values, 
     obtained in the case of a standard model nebula (\cite{C93}) and using dust simulations, are given here as an indication for the timescale/captured mass to expect from such vortices.
\subsection{Accretion Through the Dead Zone}
These simulations do not allow to follow the evolution of the vortices with enough precision.
     Therefore we focus our simulation box on either of the boundaries by putting it at the radius of $r = 1$ AU. 
     This allows us to have a much better resolution at the location of the vortices.
\begin{figure}[htbp]
\centering
\resizebox{\hsize}{!}{\includegraphics{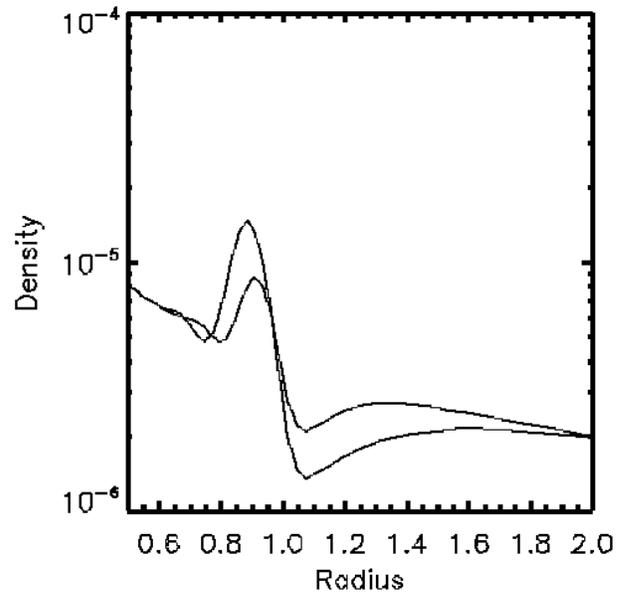}}
  \caption{This figure shows the density profile at $t=100$ (formation
     of the vortices) and 400 years : this shows that the spiral waves,  
        seen in figure 2 to extend from the vortices, cause accretion across   
          the Dead Zone.}
\label{fig:accretion_density}
\end{figure}
Figure \ref{fig:accretion_density} shows the evolution of the density profile in the region around
      the outer edge of the Dead Zone between $t=100$ years, when the RWI appears, 
      and $t=400$ years. This shows that accretion proceeds across the Dead Zone, despite the low viscosity. This can be attributed to the spiral waves generated by the vortices, and seen in figure (\ref{fig:disk_RWI_out1}) to extend across the Dead Zone. The generation of these waves corresponds to an exchange of energy and angular momentum between the vortices and the gas of the Dead Zone. We did not attempt to quantify this accretion, since this would require reaching a steady state which we can not obtain reliably with our present simulations, for a number of reasons: first, because of the 2D nature of the code we use; second, because  on longer timescales the dynamics of the vortices will be influenced by the accumulated dust, and a  two-phase code following separately both gas and dust would be required; finally, we also note that even a weak self-gravity would significantly enhance the coupling between the vortices and the spiral waves. 
\section{Conclusion}
Although the Dead Zone is not turbulent, we have shown that it naturally results in 
conditions supporting the development of the Rossby Wave Instability.  The differential mass accretion creates a high density bump/ring at the boundaries of the Dead Zone and trigger the RWI.The vortices, in turn, survive long enough to be significant for dust accretion and enhanced planet formation. They also generate spiral waves, resulting in significant accretion across the Dead Zone. 
    
Further studies will be needed to quantify these effects. This will require, in particular, 
a gas/dust/planetesimal code able to run on long time scales, and probably also the inclusion of self-gravity.   

\begin{acknowledgements}
PV is supported by NSF grants AST-9702484, AST-0098442, NASA
grant NAG5-8428, HST grant, DOE grant DE-FG02-00ER54600, the
Laboratory for Laser Energetics.

We thanks F.Masset for the code and fruitful discussion. 
PV thanks P.Barge and the OAMP for the month long stay at Marseille
where part of this work was done. 
\end{acknowledgements}


\begin{thebibliography}{}


\bibitem[D'Alessio et al., 1998]{A98} P. D'Alessio, J. Canto, N. Calvet 
  \& S. Lizano, 1998, ApJ, {\bf 500}, 411-427.

\bibitem[Balbus \& Hawley, 1991]{BH91} Balbus, S. \& Hawley, J., 1991,
  ApJ, 376, 214.

\bibitem[Barge \& Sommeria, 1995]{BS95} Barge, P.; Sommeria, J., 1995
    A\&A 295, 1.

\bibitem[Chavanis, 2000]{Ch00} Chavanis, P. H, 2000, A\&A, 356, 1089.

\bibitem[Cuzzi et al., 1993]{C93} Cuzzi, J.N., Dobrovolskis, A.R., Champney, J.M., 1993,
Icarus, 106, 102.      

\bibitem[de la Fuente Marcos \& Barge, 2001]{FMB01} de la Fuente Marcos, C.; Barge, P., 2001,
 MNRAS, 323,601.

\bibitem[Gammie, 1996]{G96} C.F. Gammie, 1996, ApJ, {\bf 457}, 355-362.

\bibitem[Goldreich et al., 1986]{GGN86} Goldreich, P.; Goodman, J.; Narayan, R.
     MNRAS, 1986, 221,339.

\bibitem[Johansen et al., 2004]{JAB04} Johansen, A.; Andersen, A. C.; Brandenburg, A., 2004
A\&A,  417, 361.

\bibitem[Li et al., 2000]{L00} H. Li, J.M. Finn, R.V.E. Lovelace \& S.A. Colgate,  2000,
  ApJ, {\bf 533}, 1023-1034.
  
\bibitem[Li et al.,2001]{L01} H. Li, S.A. Colgate, B. Wendroff \& R. Liska,  2001,
  ApJ, {\bf 551}, 874.

\bibitem[Lovelace \& Hohlfeld, 1978]{L78} Lovelace, R. V. E., \& Hohlfeld, R. G. 1978, ApJ, 221, 51.

\bibitem[Lovelace et al., 1999]{L99} R.V.E. Lovelace, H. Li, S.A. Colgate \& A.F. Nelson, 1999,
  ApJ, {\bf 513}, 805-810.

\bibitem[Masset, 2000]{masset00}
Masset F.~S., 2000, A\&AS, 141, 165

\bibitem[Masset, 2002]{masset02}
Masset, F.~S.~2002, A\&A, 387, 605

\bibitem[Masset \& Papaloizou, 2003]{masset03}
Masset, F.~S., \& Papaloizou, J.~C.~B.~2003,
accepted for publication in ApJ,
(astro-ph/0301171)
 
\bibitem[Matsumura \& Pudritz, 2005]{MP05} S. Matsumura \& R.E. Pudritz, 2005, ApJ, {\bf 618}, L137-L140.

\bibitem[Papaloizou \& Lin, 1989]{PL89} Papaloizou, J. C. B.; Lin, D. N. C.,
    ApJ, 1989, 344, 645.

\bibitem[Papaloizou \& Pringle, 1985]{PP85} Papaloizou, J. C. B.; Pringle, J. E., MNRAS, 1985,
213, 799.

\bibitem[Reyes-Ruiz, 2001] {R01}M.Reyes-Ruiz, 2001, ApJ, {\bf 547}, 465-474.

\bibitem[Shakura \& Sunyaev, 1973]{SS73} Shakura \& Sunyaev XXX

\bibitem[Stone \& Norman, 1992]{stone}
Stone, J.~M., \& Norman, M.~L.~1992, ApJS, 80, 753

\end{thebibliography}
\end{document}